\documentclass{openjournal}

\usepackage{xcolor}
\usepackage{textgreek}
\usepackage[utf8]{inputenc}
\usepackage[english]{babel}

\usepackage{hyperref}
\hypersetup{
    unicode, 
    colorlinks=true,
    linkcolor=linkcolor,
    citecolor=linkcolor,
    filecolor=linkcolor,
    urlcolor=linkcolor,
}
\usepackage{color,colortbl}
\definecolor{linkcolor}{rgb}{0.0,0.3,0.5}
\usepackage{tensind}
\tensordelimiter{?}
\DeclareGraphicsExtensions{.bmp,.png,.jpg,.pdf}
\usepackage{verbatim}
\usepackage[normalem]{ulem}
\usepackage{orcidlink}
\usepackage{soul}

\newcommand{\chandra}{{\em Chandra}}
\newcommand{\suzaku}{{\em Suzaku}}
\newcommand{\xrism}{{\em XRISM}}
\newcommand{\resolve}{{\em Resolve}}

\newcommand{\athena}{{\em NewAthena}}
\newcommand{\mach}{{\mathcal{M}_\mathrm{3D}}}
\newcommand{\vbulk}{{V_\mathrm{bulk}}}
\newcommand{\vdisp}{{V_\mathrm{disp}}}

\usepackage{soul}
\usepackage{amsmath}
\usepackage{xfrac}
\usepackage{capt-of}%

\graphicspath{ {./figs/} }

\begin{document}
\title{The major cluster merger in Abell 2034 as seen by \xrism{}: strong turbulence and spectral anomalies?}

\author{Annie Heinrich\orcidlink{0000-0002-7726-4202}$^{1}$}
\email{amheinrich@uchicago.edu}
\author{Irina Zhuravleva\orcidlink{0000-0001-7630-8085}$^{1}$}
\author{Eugene Churazov\orcidlink{0000-0002-0322-884X}$^{2,3}$}
\author{Congyao Zhang\orcidlink{0000-0001-5888-7052}$^{4,1}$}
\author{Daniele Rogantini\orcidlink{0000-0002-5359-9497}$^{1}$}
\author{Hannah McCall\orcidlink{0000-0003-3537-3491}$^{1}$}
\author{Reinout J. van Weeren\orcidlink{0000-0002-0587-1660}$^{5}$}
\author{William R. Forman\orcidlink{0000-0002-9478-1682}$^{6}$}

\affiliation{$^{1}$University of Chicago, Department of Astronomy \& Astrophysics, Chicago IL 60637, USA}
\affiliation{$^{2}$Max Planck Institute for Astrophysics, D-85741 Garching, Germany}
\affiliation{$^{3}$Space Research Institute (IKI), Moscow 117997, Russia}
\affiliation{$^{4}$Masaryk University, Department of Theoretical Physics \& Astrophysics, Brno 61137, Czechia}
\affiliation{$^{5}$Leiden Observatory, Leiden University, PO Box 9513, 2300 RA Leiden, The Netherlands}
\affiliation{$^{6}$Center for Astrophysics, Harvard and Smithsonian, Cambridge MA 02138, USA}

\begin{abstract}
    \xrism{} observations to date have shown that gas kinetic pressures in the intracluster medium (ICM) tend towards the low end of predictions from cosmological simulations. 
    Here, we present a \xrism{} observation of the merging cluster Abell 2034, which exhibits the broadest emission lines yet observed in a galaxy cluster.
    We measure a velocity dispersion of $\sim 470$ km s$^{-1}$, corresponding to a kinetic pressure fraction of $\sim15\%$.
    This places A2034 at or above the high end of the theoretical predictions for similar-mass clusters.
    This large velocity dispersion may reflect Mach $\sim0.5$ turbulence in the ICM and/or result from a core disruption driven by the ongoing head-on merger.
    We also detect a $\sim 380$ km/s gas bulk velocity gradient along the merger axis with an opposite sign to the galaxy velocity gradient, indicating a decoupling of the cluster galaxies (and dark matter) from the ICM.
    Finally, we report tentative evidence of several spectral anomalies, including a suppressed Fe He$\alpha$-$z$ line, an enhanced Fe Ly$\alpha$-2 line, and a potential absorption feature at $\sim 8.7$ keV.
    The first two features may be explained by the combination of a {multi-component} ICM and a non-equilibrium ionization state in the wake of a merger shock.
    Deeper \xrism{} observations of this cluster are required to confirm these features.
    This work highlights the importance of kinematic measurements across a large sample of merging clusters as well as the need for deep \xrism{} observations to unveil more exotic physics in the ICM.
\end{abstract}

\begin{keywords}
    {Galaxy clusters, Intracluster medium, High energy astrophysics, Spectroscopy}
\end{keywords}

\maketitle

\section{Introduction}
\label{sec:intro}

\setcounter{footnote}{0}

\xrism{} has now provided direct measurements of the kinematic properties of the intracluster medium (ICM) in approximately a dozen galaxy clusters, including several merging clusters.
\xrism{} measured the velocity structure function in the Coma cluster \citep{xrism_coma,eckert2025,vazza2026}, constrained the merger geometry in {A1914, A3667, A754, and Coma} \citep{heinrich2025,omiya_a3667,omiya-a754,zhang2026,gatuzz2026}, and detected gas-sloshing-driven gas motions in A2319 and several relaxed clusters \citep[e.g.,][]{xrism-a2319,xrism-centaurus,xrism_a2029,xrism-perseus,zhang-perseus,bellomi2025}.
{One common feature of all of these merging clusters is relatively narrow emission lines, indicating kinetic pressure fractions of $\sim 10\%$ in A3667 and $\lesssim5\%$ in all other clusters in this sample.
In addition, \xrism{} has now identified several ``exceptionally relaxed'' clusters with kinetic pressure fractions of $\lesssim 1\%$ in regions without AGN feedback \citep[A2029, A1795, and A3571,][]{xrism_a2029,sarkar2026,mccall2026}.}

A recent comparison between the observed kinematics and cosmological simulations indicated that, on average, the measured kinetic pressure fraction lies below or at the lower end of the predicted distribution \citep{xrism-sims}.
This trend is observed both inside cool-cores---where gas motions are affected by active galactic nucleus (AGN) feedback---and in regions of clusters where mergers dominate.
In the former, the discrepancy likely arises from the complexity of modeling feedback physics in simulations.
The cause of the discrepancy in the latter regions remains unclear.
It is plausible that the current sample of \xrism{}-observed clusters does not statistically represent the overall population of merging clusters in the Universe.
Expanding this sample is therefore essential to understand the cause of this discrepancy.

In this paper we present observations of another merging cluster, Abell 2034.
This cluster is of a similar mass to Coma, A1914, and A2319 \citep[$M_\mathrm{200}\approx10^{15}M_\odot$,][]{owers2013} yet exhibits broader emission lines that indicate a higher velocity dispersion and kinetic pressure fraction.
A2034 is undergoing a plane-of-sky, head-on, major merger \citep[mass ratio $\sim2.4$, ][]{monteiro-oliveira2018,moura2021,finner2025} between the two main subclusters A2034S and A2034N.
This is a similar merger configuration to the famous Bullet cluster \citep{markevitch2002}; however, A2034 is distinct in that it lacks the eponymous ``bullet'' (a cool-core remnant from the infalling cluster).
This suggests that the core of the infalling subcluster (A2034N) has been disrupted either by a previous merger or the ongoing one.
A surface brightness discontinuity is observed in the X-ray north of the cluster center.
It was originally classified as a cold front \citep{kempner2003} but was more recently determined to be a shock with Mach number $\mathcal{M}\sim1.6$ \citep{owers2013}.
A2034 also includes a number of diffuse radio sources, including a radio halo that covers the cluster's X-ray peak and a pair of radio lobes associated with the southern brightest cluster galaxy (BCG) that have been reaccelerated by a shock passage \citep[e.g.,][]{kempner2001,vanweeren2011,shimwell2016}.

This paper is organized as follows.
In Section \ref{sec:methods} we describe the data reduction and spectral modeling procedure.
In Section \ref{sec:results} we report the measured bulk velocities/velocity dispersions and their implications for the state of the merger.
We then compare the measured velocities to other \xrism{}-observed clusters and discuss anomalies in the spectrum that may indicate a departure from ionization equilibrium and a possible absorption feature (Section \ref{sec:discuss}).
{Throughout this paper, we assume a flat $\Lambda$CDM cosmology with $h=0.7$ and $\Omega_m=0.3$.}
At the A2034 redshift of $z\approx 0.1132$ \citep{monteiro-oliveira2018}, $1'$ corresponds to $\sim 125$ kpc.
Therefore, the $3'\times3'$ \resolve{} array covers a $\sim 375\times375$ kpc region.
All uncertainties are $1\sigma$ unless stated otherwise.

\section{Methods}
\label{sec:methods}
\xrism{} observed A2034 (J2000: RA=$227.5489^\circ$, DEC=$33.5105^\circ$) for a total of $\sim 127$ ks in June of 2025 (ObsID 201092010).
The observation was reprocessed using \textsc{heasoft} version 6.36, applying the most up-to-date calibration (\resolve{} \textsc{CalDB} 20250915).
The X-ray event file was filtered to include only high-resolution primary events, which have a spectral resolution of $\sim4.5$ eV \citep[see the \xrism{} Quick Start Guide\footnote{\url{https://heasarc.gsfc.nasa.gov/docs/xrism/analysis/quickstart/}} and][]{xrism_firstlight}.
We also removed pixel 27 \citep[which has been shown to have unstable gain,][]{xrism_firstlight} and pixel 7 (which was experiencing excess broadband noise during this observation\footnote{\url{https://heasarc.gsfc.nasa.gov/FTP/xrism/postlaunch/gainreports/2/201092010\_resolve\_energy\_scale\_report.pdf}}) from the event file.

\begin{figure}
    \centering
    \includegraphics[width=\linewidth]{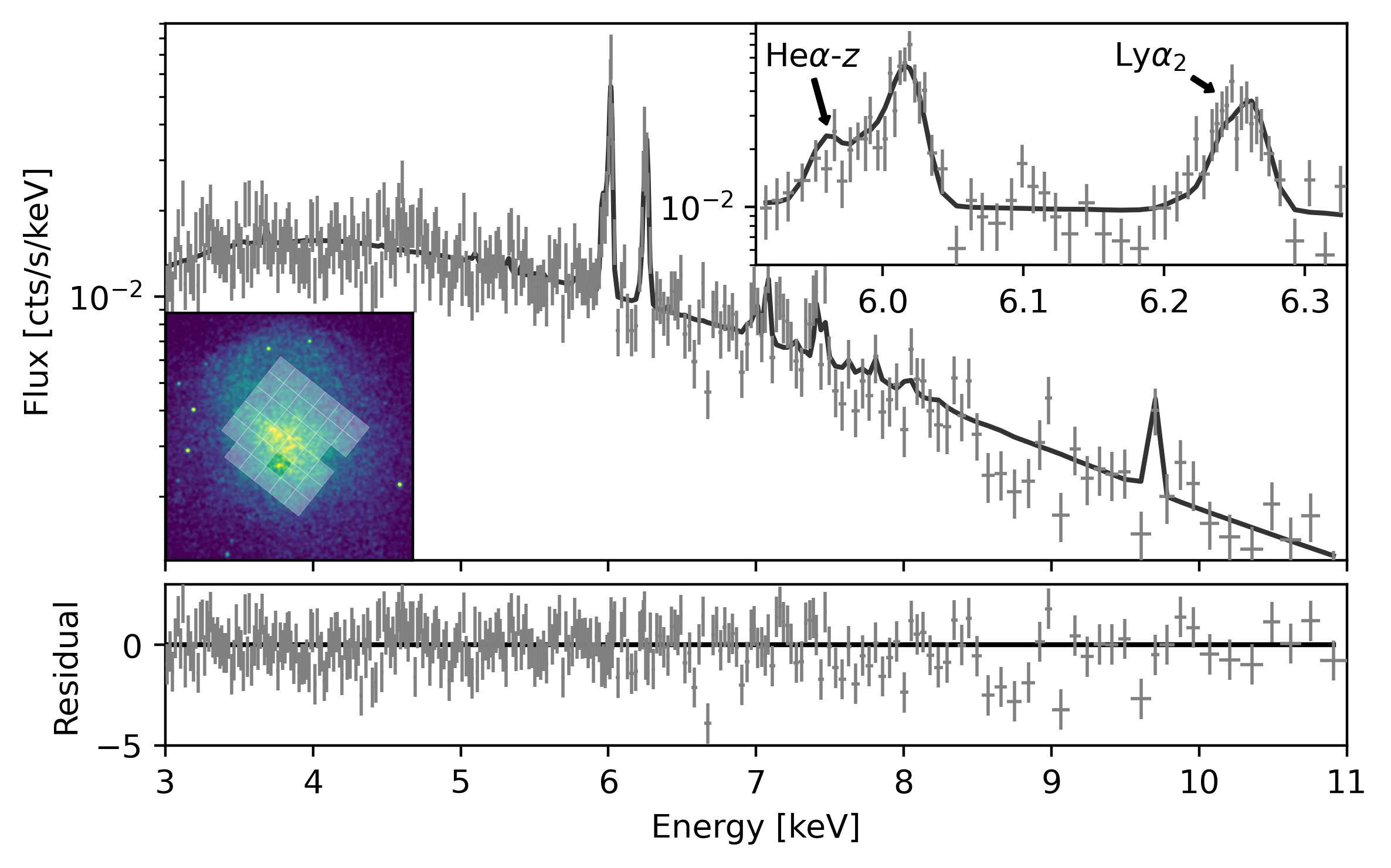}
    \caption{Full-array \resolve{} spectrum of A2034 (grey crosses) with best-fit one-component \texttt{bapec} model (black curve) and residuals. {The top right inset shows} a zoom-in on the prominent Fe He$\alpha$ and Ly$\alpha$ lines. The He$\alpha$-$z$ and Ly$\alpha_2$ residuals (Section \ref{sec:nei}) are labeled. The full spectrum is binned to $5\sigma$ significance, while the zoom-in is binned to $3\sigma$. The {bottom left} inset shows the 2-8 keV \chandra{} image of A2034 with the \resolve{} array overlaid in white. The BCGs of A2034N and A2034S are marked with stars.}
    \label{fig:spectrum}
\end{figure}

We extracted spectra from the full \resolve{} array and from the northern/southern halves of the array.
Large size response matrix files (RMFs, which are sufficient for modeling the spectra in the 3-11 keV range) were created for each spectrum using the \texttt{rslmkrmf} script.
The full-array spectrum is plotted in Figure \ref{fig:spectrum}.
We calculated effective area curves (i.e. auxiliary response files, ARFs) by running ray-tracing simulations via the \texttt{xaarfgen} script in image mode.
{Initial photon positions were determined by an exposure-corrected 2-8 keV \chandra{} image of A2034 (see Appendix \ref{app:leff}).}
As $\gtrsim99\%$ of the X-rays that reach the \resolve{} array originate within a 3'-radius circle around the FoV, we masked the image outside of this region.
We created two ARFs for the northern and southern spectra corresponding to the northern and southern halves of the {$3'\times3'$} region.
These spectra were then modeled as a combination of the emission from both regions, which accounts for spatial-spectral mixing \citep[SSM, see Appendix B of][]{heinrich2025}.

To model the contribution to the spectra from the non-X-ray background (NXB), we extracted full-array NXB spectra from the \resolve{} night-Earth database via \texttt{rslnxbgen}, applying the same filtering to the NXB spectra as the on-source spectra.
These full-array NXB spectra were fit using the \citet{kilbourne2018} model, which contains a powerlaw continuum and 8 Gaussian emission lines, using a diagonal RMF and no ARF.
All NXB parameters were fixed prior to fitting the source spectrum.
For the sub-array regions, the best-fit model was then scaled according to the number of pixels used in each on-source spectrum.

Each spectrum was fit with the NXB and ICM models in the 3-11 keV band using \textsc{xspec} version 12.15.1 \citep{arnaud1996} with Cash statistics \citep{cash1979}.
The ICM was modeled as an absorbed collisionally-ionized equilibrium (CIE) plasma (\texttt{tbabs}$\cdot$\texttt{bapec}) using \textsc{AtomDB} version 3.1.3 \citep{foster2013}.
{We additionally tested AtomDB version 3.0.9 and the 5-7 keV band, which result in no significant deviations in our findings.}
The Hydrogen column density was fixed to $1.6 \cdot10^{20}$ cm$^{-2}$ \citep{willingale2013}.
We ignored the spectrum between 8.5 and 8.9 keV due to the presence of a potential absorption-like feature, discussed in Section \ref{sec:absorp}.
The free parameters of the model are the plasma temperature, abundance {of heavy elements} \citep[{relative to} proto-solar abundances from][]{lodders2009}, redshift, velocity dispersion, and normalization.
We additionally experimented with a \texttt{bvapec} model with free Fe and Ni abundances. 
All elements lighter than Fe were tied to the Fe abundance, as emission lines from these elements are very dim due to the high plasma temperature.
The Ni abundance is measurable; however, the significance is low ($<3\sigma$) and always consistent with the Fe abundance within $1\sigma$.
Thus, using a \texttt{bapec} model is sufficient for this data.

\begin{figure}
    \centering
    \includegraphics[width=\linewidth]{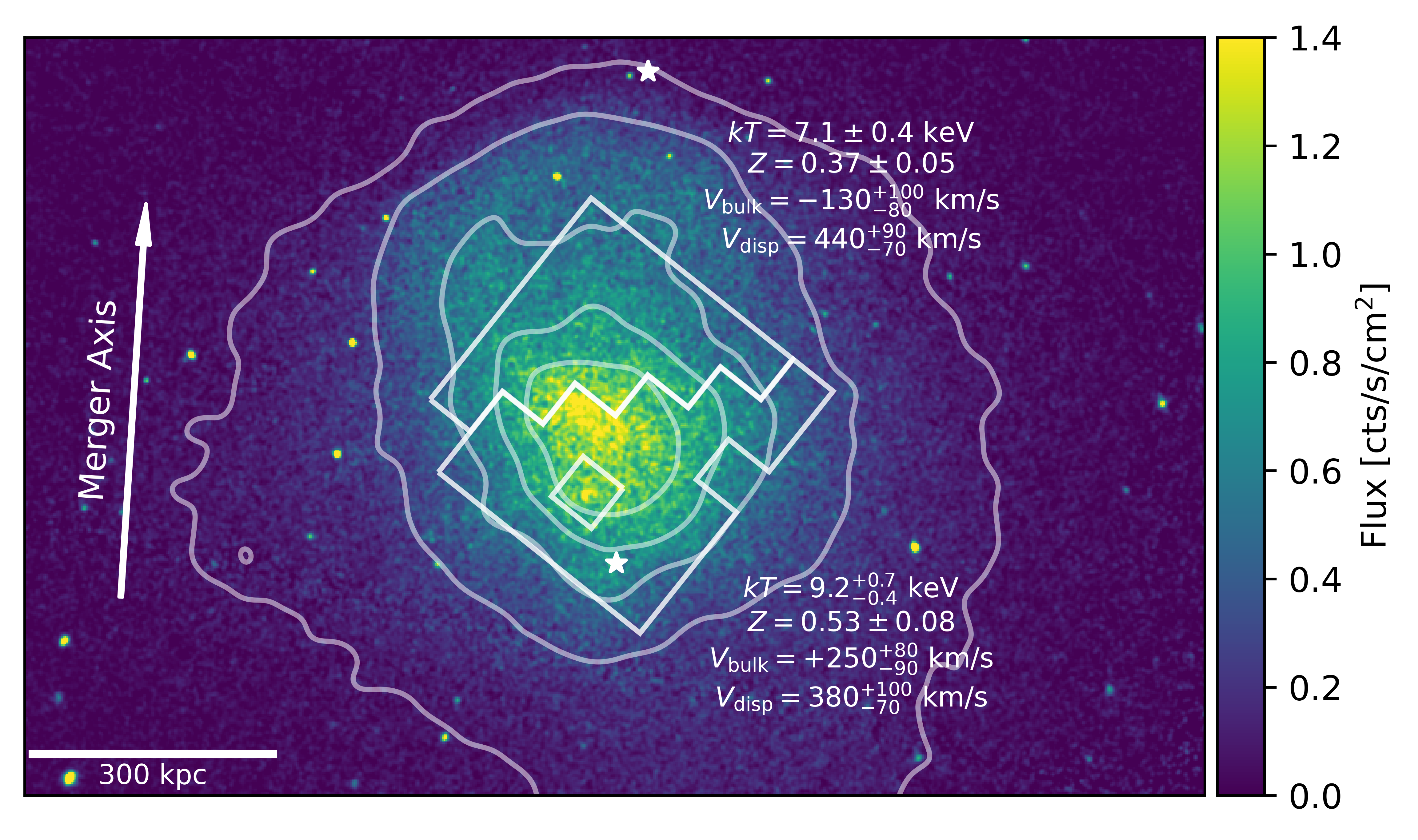}
    \caption{{\chandra{} 0.5-8.0 keV image of A2034, with X-ray surface brightness contours overlaid.
    The north and south regions are outlined in white, and the best-fit parameters for each region are listed. The BCGs of A2034N and A2034S are marked with stars. The merger axis is labeled.}}
    \label{fig:chandra}
\end{figure}

\section{Results}
\label{sec:results}
\begin{figure}
    \centering
    \includegraphics[width=\linewidth]{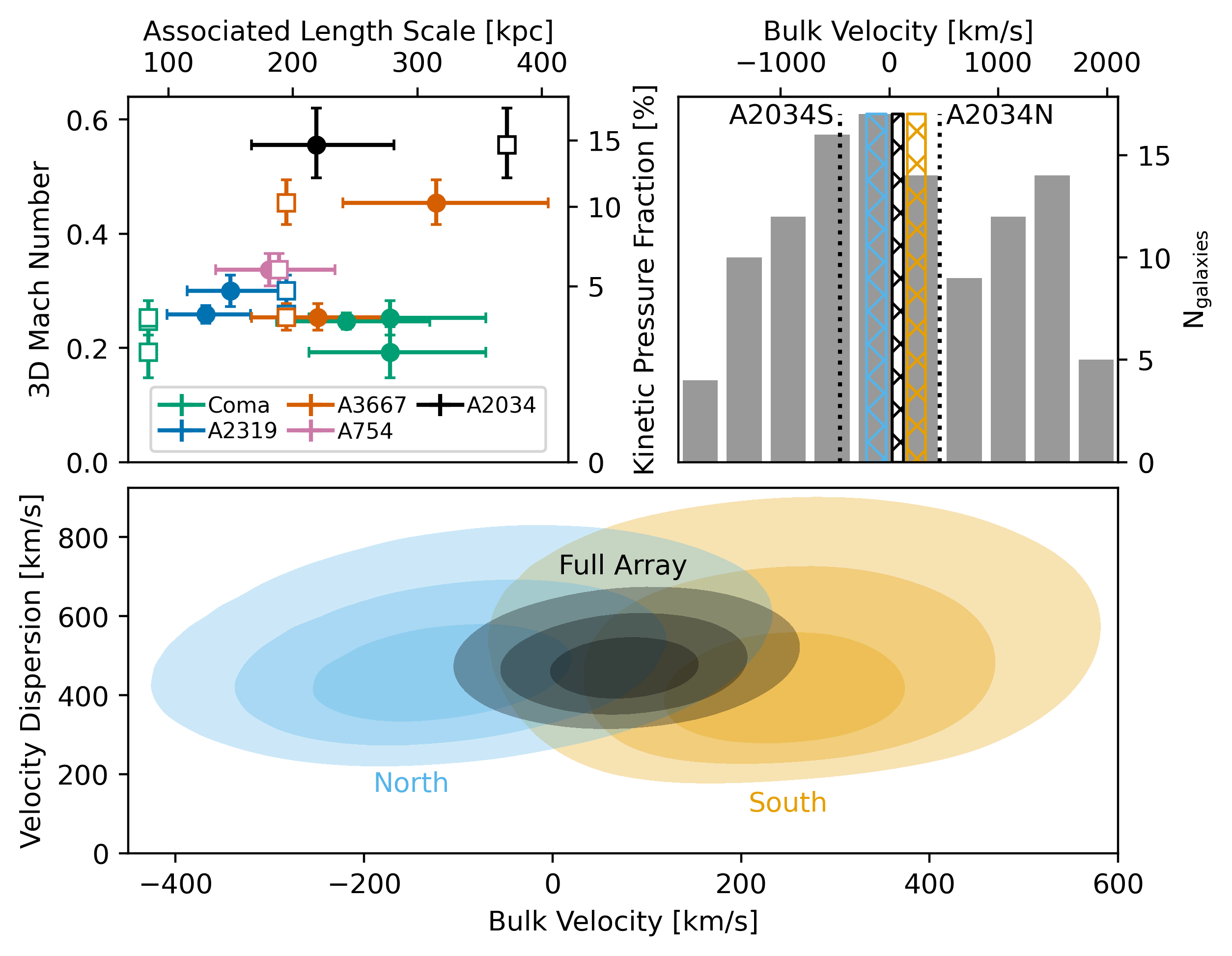}
    \caption{\textbf{Top Left:} 3D turbulent Mach number ($\mach$) and kinetic pressure fraction for A2034 (black) and other \xrism{}-observed merging clusters \citep[colors,][]{xrism_coma,xrism-a2319,gatuzz2026,omiya_a3667,omiya-a754} plotted against the associated length scale. Length scales are derived from the FoV size (unfilled squares) and the effective length scale along the LoS (filled circles). Mach numbers are calculated from the velocity dispersion measured in the full \resolve{} array with a single component CIE model. \textbf{Top Right:} A2034 member galaxy velocity histogram (grey bars) from \citet{monteiro-oliveira2018} versus the bulk velocity of the ICM measured in the full array (black hatched region) and in the north/south regions (blue/orange hatched regions). Velocities of the BCGs of A2034 are plotted as dotted black lines. \textbf{Bottom:} 1, 2, and 3$\sigma$ constraints on bulk velocity and velocity dispersion in the full array (black) and north/south regions (blue/orange) of A2034.}
    \label{fig:velcomp}
\end{figure}

The \resolve{} spectrum of A2034 is reasonably well-described by a hot CIE plasma model ($kT=8.1\pm0.4$ keV) with moderate metallicity ($Z/Z_{\odot} = 0.45 \pm 0.04 $), consistent with previous measurements from \chandra{} \citep{kempner2003,owers2013}.
The velocity dispersion of the plasma is $\vdisp=470 \pm50$ km/s---higher than has been observed in other merging clusters \citep[e.g.,][]{xrism_coma,xrism-a2319,omiya_a3667,omiya-a754}.
Meanwhile, the gas bulk velocity \citep[relative to the cluster member galaxy mean redshift of 0.1132,][]{monteiro-oliveira2018} is moderate, moving away from the observer at $\vbulk=+{70}_{-60}^{+50}$ km/s.
The full-array spectrum and best-fit model are shown in Figure \ref{fig:spectrum}, and the 1/2/3$\sigma$ constraints on the bulk velocity/velocity dispersion are plotted in black in the lower panel of Figure \ref{fig:velcomp}.

This high velocity dispersion may indicate the presence of rather strong, approaching transonic-level merger-driven turbulence in A2034, with $\mach=\sqrt{3}\vdisp/c_s=0.56\pm{0.06}$, where $c_s$ is the speed of sound and we assume the velocity dispersion reflects isotropic random motions.
{However, a superposition of multiple bulk velocity components may also contribute to the high velocity dispersion.}
{We test this scenario} by fitting the \resolve{} spectrum with a two-temperature, two-bulk-velocity model.
Velocity dispersions and metallicities must be tied between the two components so that the model converges on a physically reasonable solution.
We find a best-fit two-component model with temperatures of $11 \pm 2$ and $6.7_{-0.7}^{+0.5}$ keV and bulk velocities of $+500 \pm 300$ and $-100 \pm 100$ km/s, respectively.
Despite the second component, the velocity dispersion remains {high ($390 \pm 70$ km/s or $\mach \approx0.4-0.5$) compared to other \xrism{}-observed mergers}.
{The two-component model moderately improves the C-stat of the fit, however the Bayesian information criterion ($\text{BIC}=k \ln(n)+\text{C-stat}$, where $k$ is the number of free parameters in the model and $n$ is the number of spectral bins) favors the simpler one-component model by $\Delta \mathrm{BIC}\approx 16$.
The best-fit two-component model is plotted in Figure \ref{fig:2comp}.}

These two ICM components may be associated with the two main subclusters of A2034, which are merging along the north-south axis.
To test the relationship between the gas and galaxy/dark matter components, we split the \resolve{} FoV into north and south halves (along the merger axis) and fit each with a single \texttt{bapec} model while correcting for SSM.
By splitting the FoV, we are able to measure the temperature, metallicity, bulk velocity and velocity dispersion in both regions.

We find ICM parameters similar to the two-component model described above.
The southern ICM is hot (${9.2}_{-0.4}^{+0.7}$ keV) and redshifted ($\vbulk=+{250}_{-90}^{+80}$ km/s) while the northern ICM is cooler ($7.1 \pm 0.4$~keV) and blueshifted (${-130}_{-80}^{+100}$ km/s).
This temperature distribution is consistent with \chandra{} measurements \citep{owers2013}, which shows hotter gas reaching up to $\sim12$ keV on the southern edge of the \resolve{} FoV.
There is also an indication that the metallicity is higher in the southern region ($Z=0.53 \pm0.08$) than the northern ($Z=0.37\pm0.05$); however, the difference is $<2\sigma$.
{The velocity dispersions in the south and north regions are $ {380}_{-70}^{+100}$ and ${440}_{-70}^{+90}$ km/s, consistent with the measurement in the full-FoV.}
The bulk velocity/velocity dispersion constraints from the northern and southern regions are shown in Figure \ref{fig:velcomp} in blue and orange shaded regions, respectively.
We check the validity of these results by also splitting the \resolve{} array into east and west halves, finding no significant difference {between these regions}.

The measured bulk velocities in the northern and southern regions reveal a bulk velocity gradient of $\Delta \vbulk \approx 380 \pm130$ km/s.
This gradient partially contributes to the high velocity dispersion measured in the full FoV.
To remove the impact of the gradient, we subtract it from the velocity dispersion to find $V'_{\rm disp}=\sqrt{\vdisp^2-\Delta \vbulk^2/4} \approx 430 \pm 60$ km/s, corresponding to $\mach=0.52\pm0.07$.

\cite{monteiro-oliveira2018} finds BCG velocities for A2034N and A2034S of $\sim+500$ and $-500$ km/s relative to the mean velocity of cluster member galaxies.
The gas is therefore moving slower with respect to the cluster mean than the galaxies, which is expected given the collisional nature of the ICM.
The bulk velocities found in the full-FoV (black) and north/south regions (blue/orange) are shown as hatched regions in the top right panel of Figure \ref{fig:velcomp} in comparison to the BCG velocities (black dotted lines) and other cluster member galaxy redshifts (grey histogram).

{We note that the northern/southern ICM is traveling in the opposite direction along the LoS as the northern/southern galaxies (i.e., the northern gas is approaching us while the northern galaxies are receding and vice-versa).}
This indicates that the southern gas was likely originally associated with the northern galaxies/dark matter and vice-versa.
This demonstrates a decoupling of the collisional gas from the {collisionless} galaxies (and, by association, dark matter).
{This decoupling of the ICM and galaxies is expected during a head-on merger, and was also recently identified in MACSJ0018.5+1626 using kinetic Sunyaev–Zel’dovich measurements \citep{silich2024}.
Given this velocity structure, we can be certain that the two subclusters in A2034 have undergone pericenter passage as the dark matter/galaxies have passed each other but the gas has not.
A recent pericenter passage is also supported by the very elongated morphology of the A2034S BCG \citep{monteiro-oliveira2018}.}

Diverging bulk motions along the LoS may also contribute to $\vdisp$.
As discussed {in Section \ref{sec:intro}}, A2034 may be considered a Bullet-like merger without the bullet.
It has been proposed that the core of the infalling subcluster (A2034N) may have been disrupted by the impact between the two main ICM components in this merger \citep{owers2013}.
{In this scenario, the two gas components will spread outwards, driving gas motions perpendicular to the merger axis (i.e., along the east-west axis and the LoS).
This ``gas squashing'' will result in a distribution of diverging bulk motions that enhances the velocity dispersion.}
The high $\vdisp$ measured by \resolve{} therefore suggests a core disruption has taken place.
However, we cannot rule out the possibility that no compact core existed in either subcluster prior to the ongoing merger.
Distinguishing between these possibilities may be possible with further \xrism{} observations (which will allow for velocity measurement in smaller regions) or in the future with \athena{} {\citep{cruise2024}}.

\section{Discussion}
\label{sec:discuss}

\subsection{Comparison to other merging clusters and numerical simulations}
\label{sec:velocities}

These {kinematic} results raise the possibility of A2034 housing significantly stronger ICM turbulence {than in} other merging clusters {observed with \xrism{} so far}, {such as} Coma \citep{xrism_coma,gatuzz2026}, A2319 \citep{xrism-a2319}, A3667 \citep{omiya_a3667}, and A754 \citep{omiya-a754}.
We demonstrate this by comparing $\mach$ and the kinetic pressure fraction $\displaystyle{P_\mathrm{kin}/P_\mathrm{tot}= \left( 1+3\gamma^{-1}\mach^{-2}\right)^{-1}}$ measured in each of these clusters, where the adiabatic index $\gamma=5/3$.
{For consistency,} velocity dispersions and temperatures are taken from single-component \texttt{bapec} models of the full-array spectra for each target. 

These measurements are plotted against the length scale associated with the measured gas motions in Figure \ref{fig:velcomp} {(top left panel)}.
This associated scale is the size of the region that dominates the emission observed by \resolve{}.
Defining this scale is nontrivial. 
Here we use two approaches: either the size of the region observed in the plane-of-sky or the size of the region along the LoS that contributes the majority of X-rays to the spectrum, which we call the effective length scale $\ell_{\rm eff}$.
The former scale is defined as the size of the \resolve{} FoV ($3'$).
We note that, due to the \xrism{} PSF, some photons are collected from outside of the FoV.
However, the spectrum is dominated by emission from a region the size of or smaller than the FoV.
For instance, in A2034 $\gtrsim 50\%$ of the X-rays observed by \resolve{} originate from a circular region with a diameter of $2.2'$ ($\sim 270$ kpc) and $\gtrsim 75\%$ of the X-rays originate from within the FoV.
The calculation of $\ell_{\rm eff}$ is detailed in Appendix \ref{app:leff} \citep[see also][]{zhuravleva2012,xrism-perseus}.
We note that the gas motions in each of these clusters may be injected on scales smaller than either scale presented here.
However, cluster mergers are expected to drive gas motions on {scales of a few 100s kpc or larger} \citep[e.g.,][]{vazza2018,miniati2015}; {thus, the measured $\vdisp$ is likely not dominated by small-scale motions}.

The size of the \resolve{} FoV is plotted in Figure \ref{fig:velcomp} as an unfilled square and $\ell_{\rm eff}$ as a filled circle.
The length scales associated with the gas motions in A2034 are comparable to or slightly larger than the scales probed in other clusters.
Despite the similarity in scale, A2034 has a higher kinetic pressure fraction than any other cluster in this sample ($\sim 15\pm3\%$), with only A3667 coming close ($\sim10\pm2\%$).
We note that sub-FoV fits in other clusters can yield regions with turbulent Mach numbers similar to A2034, e.g. $\mach \approx 0.7 \pm 0.2$ in A3667 and $\mach\approx0.6 \pm0.2$ in A754 \citep{omiya_a3667,omiya-a754}.

\citet{xrism-sims} compared the velocity dispersions measured by \xrism{} in several clusters to {tailored} predictions from a number of cosmological simulations.
Two non-cool-core clusters were included in that sample, A2319 and the Coma cluster, both of which were found to have kinetic pressure fractions of $\sim 3\%$.
{This is on the low end of the {prediction} from simulated clusters ($P_\mathrm{kin}/P_\mathrm{tot} \sim3-11\%$).}
A2034 is similar in mass\footnote{We note that the $M_{200}$ found in A2034 ranges from $\sim 0.5-1.1 \cdot 10^{15} M_\odot$ depending on the method used \citep{monteiro-oliveira2018,finner2025}. However, given the value for $M_{500}$ found by Planck \citep{planck2013} is $>0.5\cdot10^{15}M_\odot$, we favor the higher mass estimate.} to both A2319 and Coma, so these clusters {can be directly compared}.

With $P_\mathrm{kin}/P_\mathrm{tot}\approx15\%$, A2034 is consistent with/above the upper end of predictions from cosmological simulations.
This indicates that the apparent trend of lower kinetic pressure in observed clusters may be a product of a limited sample size.
{In other words, the addition of A2034 demonstrates that merger-driven gas kinematics in the ICM can vary significantly from cluster-to-cluster.}
We note that this comparison must be verified by simulation analysis tailored to match the observed region of A2034 (about twice larger than the region observed in A2319), which is beyond the scope of this work.
However, the region covered in A2319 is itself twice the size of the region in Coma, and the kinetic pressure fraction changes minimally between the simulated analogues of these two clusters.
Therefore, region size is unlikely to significantly affect the measured kinetic pressure fraction, given that the effective length along the LoS is similar for all three clusters (Figure \ref{fig:velcomp}).

\subsection{Spectral anomalies}
\label{sec:nei}
While the one and two-component models described above provide reasonable fits to the A2034 spectrum, there are a few features that are not explained by this baseline model:
\begin{itemize}
\item 
    \textit{Ly$\alpha$ anomaly:}
    A2034 shows a brighter Fe Ly$\alpha$-2 line {than is predicted by the baseline model}.
    {This is visible as a positive residual from the baseline model (black curve) at $\sim 6.25$ keV in Figure \ref{fig:nei-abs}, detected at $\sim2\sigma$ significance.}
    While the value of the Ly$\alpha$-2/1 ratio for this temperature gas is expected to be $\sim 0.5$, the {observed} ratio in A2034 is higher.
    We measure the ratio by freezing the temperature, metallicity, and velocity dispersion of the single-component model and replacing the Ly$\alpha$ lines with two Gaussians.
    The width and redshift of each Gaussian are tied to the \texttt{bapec} model, while the normalizations are free.
    In the full FoV, the ratio Ly$\alpha$ ratio is ${1.3}_{-0.4}^{+0.8}$.
    A similar anomalously high Ly$\alpha$ ratio was observed in the Coma and A2029 clusters \citep{xrism_coma,xrism_a2029}; however, the difference is more prominent in A2034.
\item
    \textit{He$\alpha$-$z$ anomaly:}
    We detect {a suppression of the} Fe He$\alpha$ forbidden ($z$) line {relative to} the baseline model (Figure \ref{fig:nei-abs} at $\sim5.95$ keV).
    {In optically-thin CIE plasmas} {with typical ICM temperatures}, the forbidden-to-resonance ratio ($z/w$) of the Fe He$\alpha$ triplet is expected to be $\sim 0.3$.
    The ratio observed in A2034 is $\sim2\sigma$ lower, at ${0.16}_{-0.06}^{+0.07}$.
    This anomaly appears more strongly in the southern half of the FoV, where the $z$ line is not detected at the $1\sigma$ level.
\end{itemize}   

{Modifying the baseline model by including multiple temperature, metallicity, or velocity components cannot explain the anomalous $z/w$ line ratio, though it can partially reduce the Ly$\alpha$-2 residual (see Figure \ref{fig:2comp}).} 
Resonant scattering \citep[see, e.g.,][for reviews]{churazov2010,gu2018} is unlikely to be the cause either.
The optical depth of the {potentially optically thick} $w$-line is $\sim 0.3$ (assuming a spherically-symmetric ICM and no turbulent broadening)---too small to create a noticeable distortion due to resonant scattering.
Moreover, resonant scattering should increase the $z/w$ ratio relative to the optically-thin prediction (assuming the ICM is not extremely asymmetric along the LoS).
Resonant scattering cannot explain the anomalous Ly$\alpha$ ratio either, as the optical depth for both Ly$\alpha$ lines is $<0.1$.
{Charge exchange is also unlikely to be significant in this cluster, as this requires the presence of cold ($T\sim10^4$ K) gas, similar to the H$\alpha$ emission in many cool-core clusters \citep[e.g.,][]{ gu2023}.}

Though the significance of these features is low ($\sim 2\sigma$, individually), their presence, if statistically confirmed, points to some intriguing possibilities for ICM physics.
{One (rather} exotic) explanation for the observed line ratios is a departure from CIE \citep[see, e.g.,][for review]{gu2018}.
{The signatures of a} non-equilibrium ionization state has been previously {investigated via numerical simulations by \citet{akahori2012}} and reported in A754 \citep{inoue2016} and A3667 \citep{shishido2025} using \suzaku{} observations.
However, these spectra do not resolve the individual He$\alpha$ and Ly$\alpha$ lines as \xrism{} does.
At \suzaku{}'s spectral resolution an apparent deviation from CIE could also be explained by multiple temperature/velocity components, which is reasonable to expect given the mergers ongoing in both clusters.
Notably, both A754 and A3667 have been observed with \xrism{} \citep{omiya_a3667,omiya-a754}, yielding spectra which do not show signs of a suppressed $z/w$ ratio.
However, both have residuals in the Ly$\alpha$ lines, which may indicate a secondary hot ICM component or a mixture of CIE and NEI plasmas.

To explore this idea, we consider a {model where a} merger shock heats the ICM.
In the immediate wake of the shock, the plasma temperature has risen substantially, but the gas has not yet reached ionization equilibrium, leading to specific spectral signatures \citep{mewe1978,liedahl1999}. 
In the shocked ICM condition, the collisional excitation of the $w$ line increases while the main channels leading to $z$ line emission (H-like ions recombination or Li-like ions {inner-shell} ionization) become subdominant, causing a decrease in the $z/w$ ratio \citep{churazov-nei}.

This NEI state is short lived; however, it is not unrealistic to suggest we may observe it in this cluster.
{The ionization parameter $\tau =n_e\cdot t$ (where $n_e$ is the electron density and $t$ is the ionization timescale) for an NEI ICM is expected to be {$\lesssim$ a few $10^{12}$ s cm$^{-3}$} \citep{churazov-nei}.}
For typical ICM densities of $\sim 10^{-3}$ cm$^{-3}$, the NEI state can last up to $\sim 30$ Myr.
{However, as the shock passes through the ICM more gas will be brought out of CIE, effectively trailing an NEI region behind the shock.
Therefore, finely tuned merger timing is not required to expect that we may detect emission from the NEI gas.}

If a merger shock is within the \resolve{} FoV we may expect three ICM states to contribute to the X-ray emission: a pre-shock, low-temperature ($\sim 6-7$ keV) gas in CIE; a post-shock NEI gas; and a post-shock, high temperature ($>10$ keV) gas that has returned to CIE.
We probe this configuration by modeling the full-FoV \resolve{} spectrum with a combination of three plasma components: two \texttt{bapec} components---representing the pre- and post-shock CIE gas---and a \texttt{brnei} component that represents the NEI gas in the wake of the shock.
The temperature of the pre-shock \texttt{bapec} model is tied to the initial temperature of the \texttt{brnei} model, while the post-shock \texttt{bapec} and final \texttt{brnei} temperatures are tied.
We additionally tie the abundances and velocity dispersions between each component.
The normalizations and redshifts of each component are free.

\begin{figure}
    \centering
    \includegraphics[width=\linewidth]{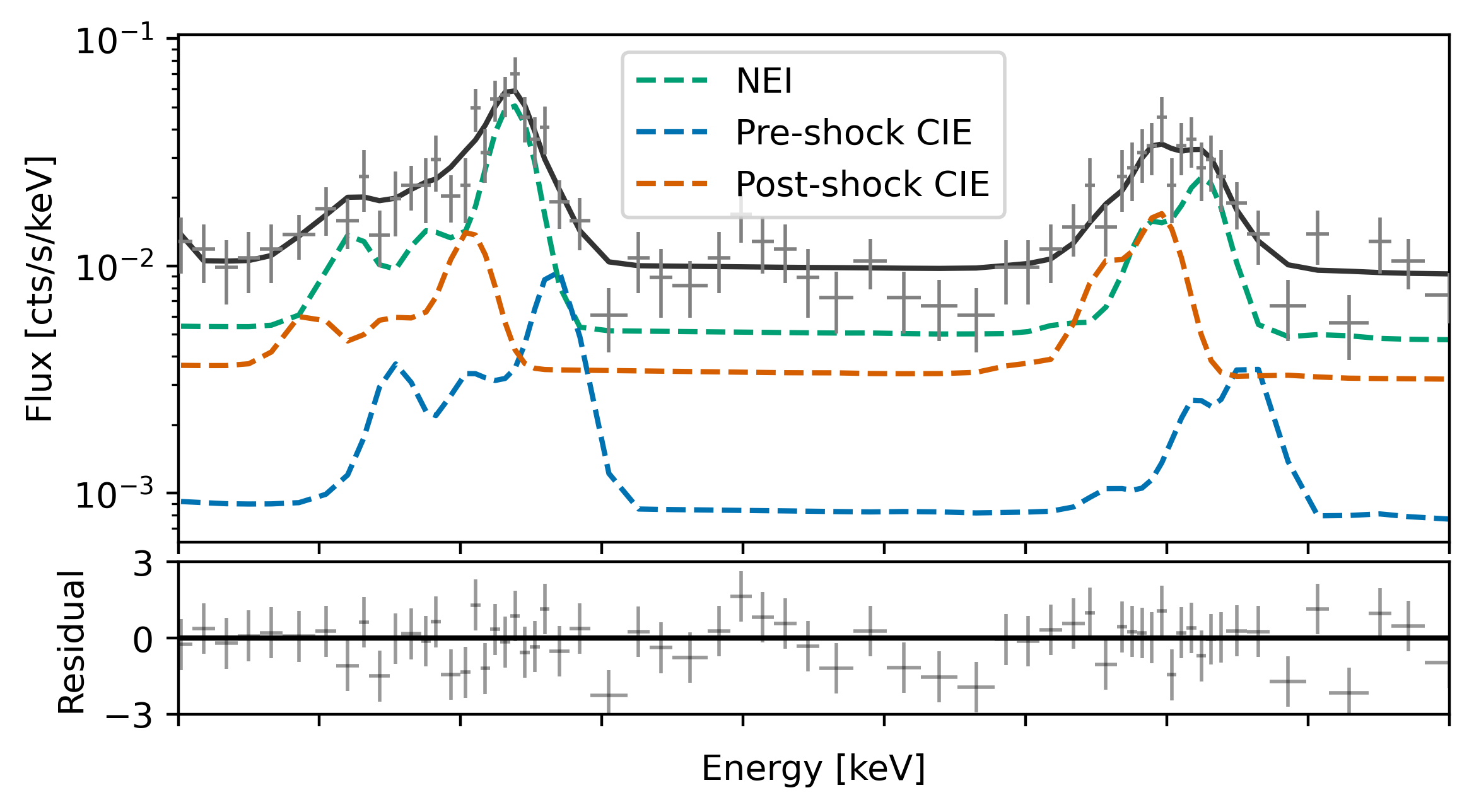}
    \caption{Full-array A2034 spectrum with best-fit CIE+NEI model (black solid curve) and its components (colored dashed curves), with residuals plotted in the lower panel. The spectrum is visually binned to $>3\sigma$ significance.
    }
    \label{fig:nei-comp}
\end{figure}

This model is fit to the full-FoV spectrum in the $5-7$ keV band, focusing on the He$\alpha$ and Ly$\alpha$ line complexes and their observed residuals.
We find a best-fit pre-shock temperature of $7_{-3}^{+1}$ keV and a post-shock temperature of $12_{-2}^{+5}$~keV, with a velocity dispersion of $300_{-60}^{+80}$ km/s.
This temperature jump is produced by a $\mathcal{M}\sim 1.8$ shock, consistent with typical shock strengths for cluster mergers.
{$\tau$ is found to be $\lesssim10^{11.6}$ s cm$^{-3}$, consistent with an NEI plasma.}
The best-fit model is plotted in Figure \ref{fig:nei-abs} in comparison to a single CIE component model and is broken down into its components in Figure \ref{fig:nei-comp}.
These figures show that the low $z/w$ ratio and high Ly$\alpha$ ratio can be explained by the NEI gas and redshifted post-shock CIE gas, respectively. 
The NEI component is the brightest of the three, followed by the post-shock CIE component, while pre-shock component is $\sim 4$ times dimmer than the others.
{Compared to the default, single-\texttt{bapec} model, this NEI model provides a better fit to the data ($\Delta\text{C-stat}\approx 7$). 
However, the added complexity of the model is disfavored according to the BIC ($\Delta\text{BIC}\approx24$ in favor of the single-\texttt{bapec} model).}
{
Fitting the $3-11$~keV spectrum with the same combination of CIE and NEI components results in a model with a bright pre-shock CIE and dim NEI component, which does not address the anomalous $z/w$ ratio.
Therefore, additional \xrism{} observations of this cluster that will increase the significance of the tentatively detected deviations are needed to confirm this solution.}

To further probe the ionization state of the ICM, we employ line ratio diagnostics that are sensitive to the excitation and ionization temperatures \citep{nevalainen2010,hitomi-temp,sarkar-a2029}.
With the current data, we are able to measure the Fe He$\alpha$ $(x+y+z)/w$ and He$\alpha$ $z$/Ly$\alpha$ line ratios, which are respectively sensitive to the excitation and ionization temperatures.
The latter ratio is $0.19\pm0.08$, consistent with an ionization temperature of $11^{+3}_{-2}$ keV in a CIE plasma.
This is similar to the post-shock temperature found in the previous {narrow-band} modeling exercise.
The former ratio is $0.31^{+0.12}_{-0.05}$, which is inconsistent with any CIE emission in the ICM temperature range.
The $(x+y+z)/w$ ratio is expected to be $\sim0.55-0.7$ for a plasma with a similar temperature to A2034, which is in $\gtrsim2\sigma$ tension with the observed ratio. 
This {further supports the} notion that the ICM in A2034 {deviates from the} ionization equilibrium.

{We note the possibility that an NEI plasma has not reached ion-electron temperature equilibrium.
In the wake of the shock, the ion temperature may be higher than the electron temperature.
If that is the case, the Fe emission lines would be thermally broadened more than expected and therefore a CIE model would overestimate the velocity dispersion based on the line widths.
Independently measuring the ion temperature requires resolving lines from multiple elements, which is not possible with the current data.}

The NEI model {we consider} {requires the presence of} a merger shock within or near the \resolve{} FoV, {yet the only reported shock in A2034 is the Mach $\sim1.8$ shock located $\sim 1'$ north of the FoV \citep{owers2013}.}
A southern shock is expected in/near the FoV based on the merger configuration \citep{moura2021}.
It is also expected to be weaker and {cover a smaller region than the northern shock, given the measured mass ratio of the merging components}. 
{Thus, the shock may be hidden by projection effects due to the bright core or a shock orientation along the LoS. }
{Further evidence for a shock in this region is given by the presence of a hot ($\sim12$ keV) region near the southern edge of the FoV \citep{owers2013}.
Finally, a radio structure just south of the FoV \citep[which is likely either a radio relic or phoenix,][]{shimwell2016} also indicates a recent shock passage.
}

\begin{figure}
    \centering
    \includegraphics[width=\linewidth]{figures/nie-abs.png}
    \caption{\textbf{Left:} Full-array A2034 spectrum comparing the predicted He$\alpha$ and Ly$\alpha$ emission lines for a CIE model (black solid curve) versus a model containing both CIE and NEI components (red dashed curve). The spectrum is visually binned to $>3\sigma$ significance.
    \textbf{Right:} The same spectrum comparing a single-component CIE model with (pink dashed curve) and without (black solid curve) a Gaussian absorption component at 8.7 keV.
    Residuals from each model are plotted below in their respective colors and offset to aid visualization.
    This spectrum is binned to $>5\sigma$ significance.
    }
    \label{fig:nei-abs}
\end{figure}

{Longer} \xrism{} observations of A2034 {are necessary} to statistically confirm the anomalous line ratios {and further explore this potential} non-equilibrium ionization state in the ICM.
To estimate the required exposure time, we produce mock \textit{Resolve} spectra using the \texttt{xspec} \texttt{fakeit} routine, based on the best-fit, full-FoV model and its accompanying response files.
By generating 100 mock spectra for various exposure times, we find the median time required to confirm the $z/w$ anomaly (the driving signature of the NEI state) at $>3\sigma$ significance is an additional $\sim360$ ks.
Essentially, a factor of $\sim4$ increase in the number of photons is needed to confirm this result.
The additional \resolve{} exposure would also yield a $\sim4.5\sigma$ detection of the anomalous Ly$\alpha$ ratio.

\subsection{Potential absorption feature}
\label{sec:absorp}

{Careful examination of the A2034 spectrum has revealed} another unique anomaly: a ``dip'' in the spectrum between $\sim$8.5 and 8.9 keV {(Figure \ref{fig:nei-abs}, the right panel)}.
The exact shape of this feature is unclear, given the low photon counts in the hard X-ray band. 
At low resolution (the spectrum shown in Figure \ref{fig:nei-abs} is {visually} binned to $>5\sigma$ significance), it can be modeled as a Gaussian absorption line (\texttt{gabs} model in \texttt{xspec}).
Modeling this feature in the full FoV spectrum, we find a detector-frame centroid of $E_\mathrm{abs}=8.70\pm0.03$ keV with a width of $\sigma_\mathrm{abs}=80\pm30$ eV and an optical depth of $\tau_\mathrm{abs}=0.8\pm0.3$. 
This feature is detected at $\sim 4\sigma$ significance.
{We also fit this feature in the north/south half-FoV spectra, finding the feature at $8.72 \pm 0.01$ ($8.69_{-0.07}^{+0.08}$) keV in the north (south) region, with an optical depth of $0.6_{-0.2}^{+0.4}$ ($3.2_{-1.6}^{+8.5}$).
The difference in optical depth is due to the feature being narrower in the north ($28_{-8}^{+9}$ eV) compared to the south ($110_{-30}^{+50}$).
When modeling the north and south ICM separately, the feature is detected at $\gtrsim2.7 \sigma$ significance in both regions.}

The presence of such an absorption feature in the spectrum of a galaxy cluster is puzzling.
To the best of our knowledge, no such features have been reported previously, and they are not expected on physical grounds unless rather unusual conditions are invoked. 
We verified that the NXB spectrum shows no emission lines in the 8-9 keV band that could have been incorrectly subtracted to produce an artificial absorption feature. 
We also confirmed with the \textit{XRISM} support team that this feature is very unlikely to be associated with any instrumental or calibration effects (private communication).

The energy and width of the feature are consistent with an ultrafast outflow (UFO) from an AGN \citep[e.g.,][]{xrism-ufo}. 
Although no AGN is detected within the detector FoV, based on the \textit{Chandra} image, there are several bright point sources near the FoV (see Figure \ref{fig:chandra}) that could, in principle, contribute to the observed spectrum as light scattered by the PSF. 
The presence of a UFO would also require a significant contribution of the AGN continuum to the spectrum.
It is possible to add an AGN component to the model described in Section \ref{sec:results}, effectively reducing the continuum emission from the ICM. 
However, this will increase the ICM metallicity to $\gtrsim0.8$, which is rather high compared to typical values of $\sim 0.3-0.5$ in the ICM and inconsistent with the metallicity measured by \chandra{} \citep{kempner2003}.
Additionally, the feature would have to absorb essentially all of the AGN continuum emission from $8.5-8.9$~keV range, meaning the UFO would need to have an extreme optical depth. 
Therefore, we consider this explanation unlikely.

Another rather exotic possibility is that the feature comes from a shock in the outskirts of A2034. 
{If such a shock (unrelated to the ongoing merger) is oriented parallel to the LoS (i.e., the shock is moving tangentially with respect to the cluster radius), it may have a significant column density.
However, this shock would be very difficult to detect in emission due to the low gas density in the cluster outskirts.}
In this case, the feature could be attributed to absorption by a high energy atomic transition {in the shocked region}.
The highest energy atomic transition expected to be strong in the ICM is the Ni Ly$\alpha$ lines, at $\sim 8.1$ keV.
For this line to reach 8.7 keV, it must be blueshifted by $\sim50\cdot10^3$ km/s (or $\sim0.16c$) relative to the cluster mean, which is unrealistic to expect in a galaxy cluster.
There are weaker lines at higher energies that would require a smaller blueshift (e.g., the Fe He$\gamma$-$\iota$ lines); however, these would be paired with stronger, lower-energy lines (e.g., He$\alpha$) that we do not see in absorption in the \resolve{} spectrum.

In conclusion, we cannot identify a physically motivated explanation for the absorption feature. 
Given its current $\sim 4\sigma$ significance, it may still be a statistical fluctuation. 
{To investigate this possibility further, we generate 100 mock \resolve{} spectra of our default ICM model (\texttt{tbabs}$\cdot$\texttt{bvapec}) with the same exposure time as this observation. 
We then fit each spectrum with a \texttt{tbabs}$\cdot$\texttt{bvapec}$\cdot$\texttt{gabs} model.
We find that $36\%$ of the mock spectra contain an ``absorption feature'' detected at the $3\sigma$ level, while only $2\%$ contain a detection of an absorption feature at the $4\sigma$ level.
Thus, the chances that this feature is a mere statistical fluctuation are low, but not non-existent.}
Deeper \textit{XRISM} observations are therefore required to confirm the feature's presence, investigate possible multi-component structure, and explore its spatial variation within the cluster.
We note that if the $4\sigma$ detection of the feature is due to mere statistical fluctuations, the best-fit ICM temperature reduces from $\sim 8.1$ keV to $7.9\pm0.3$ keV, while all other parameters are unchanged.

\section{Conclusions}
We conducted a detailed study of the high-resolution X-ray spectrum measured by \xrism{}/\resolve{} in the merging cluster A2034, revealing several unique features that could not have been discovered without \xrism{}'s spectral resolution.
The standout feature is the velocity dispersion, which is higher than in other merging clusters observed by \xrism{} so far: $\sim400-500$ km/s in the full FoV, corresponding to a 3D turbulent Mach number of $\sim 0.56$.
{As expected for a near-plane-of-sky merger, the average bulk velocity of the ICM in the full FoV is consistent with 0 km/s.} 
The main conclusions can be summarized as follows:
\begin{itemize}
    \item 
    We probe the spectrum for evidence of multiple bulk velocity components along the line-of-sight.
    We find that the combination of a hot ($kT\approx 11$ keV) component that is redshifted with respect to the cluster mean ($\vbulk\approx+500$ km/s) with a cooler blueshifted component ($kT\approx 6.5$ keV, $\vbulk\approx-100$ km/s) provides a better fit to the spectrum.
    When separating these bulk velocity components, the velocity dispersion of both components is $\sim 390$ km/s.
    \item 
    By splitting the FoV along the north-south merger axis, we show that the hot component is localized in the south and the cooler component in the north.
    The bulk velocity gradient between the two regions is $\sim380$ km/s, consistent with the near-plane-of-sky, head-on merger in A2034.
    \item    
    The southern/northern gas are moving in the opposite directions along the LoS of the southern/northern galaxies, demonstrating a decoupling between the two, which is expected in dissociative mergers.
    The gas is also moving slower compared to the BCGs of the northern/southern subclusters, consistent with the collisional nature of the ICM.
    This indicates the merger in A2034 is just past pericenter passage as the galaxies have had time to pass each other but the gas has not.
    \item 
    We discussed that diverging bulk motions along the LoS associated with the disruption of the subcluster cores during the merger (``gas squashing'') may be partially responsible for the high velocity dispersion.
    \item 
    We compare the turbulent Mach number ($\mach$) in A2034 to other merging clusters observed by \xrism{}, calculating the gas motions' associated length scale in each.
    While the length scales probed in A2034 {(i.e., a few hundred kpc)} are similar to other clusters, it has a turbulent Mach number of $\sim0.56$ and kinetic pressure fraction of $\sim15\%$, higher than any other merging cluster yet observed.
    {This kinetic pressure fraction puts A2034 on the high end of the $P_\mathrm{kin}/P_\mathrm{tot}$ distribution found in analogous clusters in cosmological simulations \citep{xrism-sims}.
    Thus, A2034 helps to balance the distribution of clusters observed by \xrism{}, which have so far skewed to low kinetic pressures.}
\end{itemize}

We additionally find several {tentative} residuals in the \resolve{} spectrum that are not {captured by} an optically thin, collisionally-ionized equilibrium plasma.
Namely, these are: a low He$\alpha$ $z/w$ ratio, a high Ly$\alpha$ 2/1 ratio, and a dip in the spectrum at $\sim 8.7$ keV indicating an unexplained absorption feature.
These features are detected at $\sim2$, 2, and $4\sigma$, respectively.

The anomalous line ratios could be explained if the ICM is in a non-equilibrium ionization state following the passage of a merger shock.
The low $z/w$ ratio, if confirmed with longer \xrism{} observations, would indicate an NEI state in the ICM, the first robustly discovered via this diagnostic criterion.
The high Ly$\alpha$ ratio can be explained by a redshifted, hot ($kT>10$ keV) ICM component.
We model the narrow-band spectrum of A2034 with a combination of NEI and CIE components, finding that a $\mathcal{M}\approx1.8$ shock that heats the gas from $\sim7$ to 12 keV can reproduce the observed He$\alpha$ and Ly$\alpha$ emission lines.
Both line ratio anomalies require additional \xrism{} exposure time to statistically confirm.

We fit the hard X-ray absorption feature with a Gaussian absorption line, finding a centroid of $\sim 8.7$ keV, width of $\sim 70$ eV, and an optical depth of $\sim 0.8$.
We do not detect significant variation in the absorption feature's spectral properties across the FoV or over the time interval of the observation. 
We speculate on the possible origin of this feature, however we cannot {find a physically-convincing} explanation without further data.

A2034 stands out from other \xrism{}-observed merging clusters due to the higher-than-previously-observed gas kinetic pressure fraction and the anomalous spectral features it produces.
This work demonstrates the need for deep \xrism{} observations of massive merging clusters---as well as {supporting} theoretical studies {aimed at} \resolve{}'s capabilities---to understand the physics that produce these and other spectral anomalies.

\section*{Acknowledgments}
AH and IZ would like to thank F. Scott Porter and Caroline Kilbourne for their helpful conversations regarding \resolve{} detector and calibration effects, as well as Yuri Ralchenko and Ildar Khabibullin for their helpful discussions on non-equilibrium ionization in the ICM.
AH and IZ acknowledge partial support from the NASA grant number 80NSSC25K7693. 
CZ acknowledges the support of the Czech Science Foundation (GACR) Junior Star grant no. GM24-10599M.

\bibliographystyle{apsrev4-1}

\bibliography{oja-vers}

\begin{appendix}
\renewcommand{\thefigure}{A\arabic{figure}}
\setcounter{figure}{0}
\renewcommand{\thetable}{A\arabic{table}}
\setcounter{table}{0}

\section{Chandra data analysis and effective length scale}
\label{app:leff}

We use exposure-corrected 2-8 keV \chandra{} images to calculate \resolve{} effective area curves and to calculate $\ell_\mathrm{eff}$ in A2034 and the other merging clusters shown in Figure \ref{fig:velcomp}.
The \chandra{} ObsIDs used are listed in Table \ref{tab:chandra}.
Each ObsID was reprocessed using the standard method \citep{vikhlinin2005} and filtered to remove flares with \textsc{CIAO} version 4.17.0 and \textsc{CalDB} version 4.12.2.
Background event files were produced using the \texttt{blanksky} and \texttt{readout\_bkg} scripts.
Exposure-corrected, background subtracted mosaic images were made using the \texttt{merge\_obs} script.
A broad-band (0.5-8.0 keV) image of A2034 is shown in Figure \ref{fig:chandra}.

\begin{table}[]
    \centering
    \begin{tabular}{c|ccc}
     Cluster & ObsIDs & $r_c$ [kpc] & $\beta$ \\
    \hline
    A2034 & 2204, 7695, 12885-6, 13192-3 & 210 & 0.64\textbf{} \\
    A754 & 6794-7, 6799, 6800, 10743 & 160 & 0.56 \\
    A2319 & \cite{storm2015} & 130 & 0.64 \\
    A3667 & 5751-3, 6292, 6295-6 &  220 & 0.60 \\
    Coma & \cite{churazov2021} & 290 & 0.73 \\
    \end{tabular}
    \caption{\chandra{} ObsIDs used to generate X-ray images, { and calculated} $\beta$-model parameters for each cluster shown in Figure \ref{fig:velcomp}. Model parameters for the Coma and A2319 clusters are taken from the listed references.}
    \label{tab:chandra}
\end{table}

The effective length $\ell_\mathrm{eff}$ is calculated by measuring a X-ray surface brightness radial profile ($S$) for each cluster and fitting it to a $\beta$-model: $\displaystyle{S\propto\left(1+(\sfrac{r}{r_c})^2\right)^{\sfrac{1}{2}-3\beta}}$, where $r$ is the projected radius from the cluster center and $r_c$ is the core radius.
Assuming spherical symmetry, the 3D X-ray emissivity ($\epsilon$) follows a three-dimensional $\beta$-model ($\displaystyle{\epsilon \propto \left( 1+(\sfrac{r}{r_c})^2+(\sfrac{z}{r_c})^2 \right)^{-3\beta}}$, where $z$ is the depth along the LoS.
We can integrate the 3D $\beta$-model along the LoS to calculate the size of the region within which the majority {(50\% of total)} of the X-ray emission originates from, which is defined as  $\ell_\mathrm{eff}$ \citep{zhuravleva2012}.
Upper/lower uncertainties on $\ell_\mathrm{eff}$ are calculated from the size of the regions that emit $60$/$40\%$ of the X-rays.

\begin{figure}
    \centering
    \includegraphics[width=\linewidth]{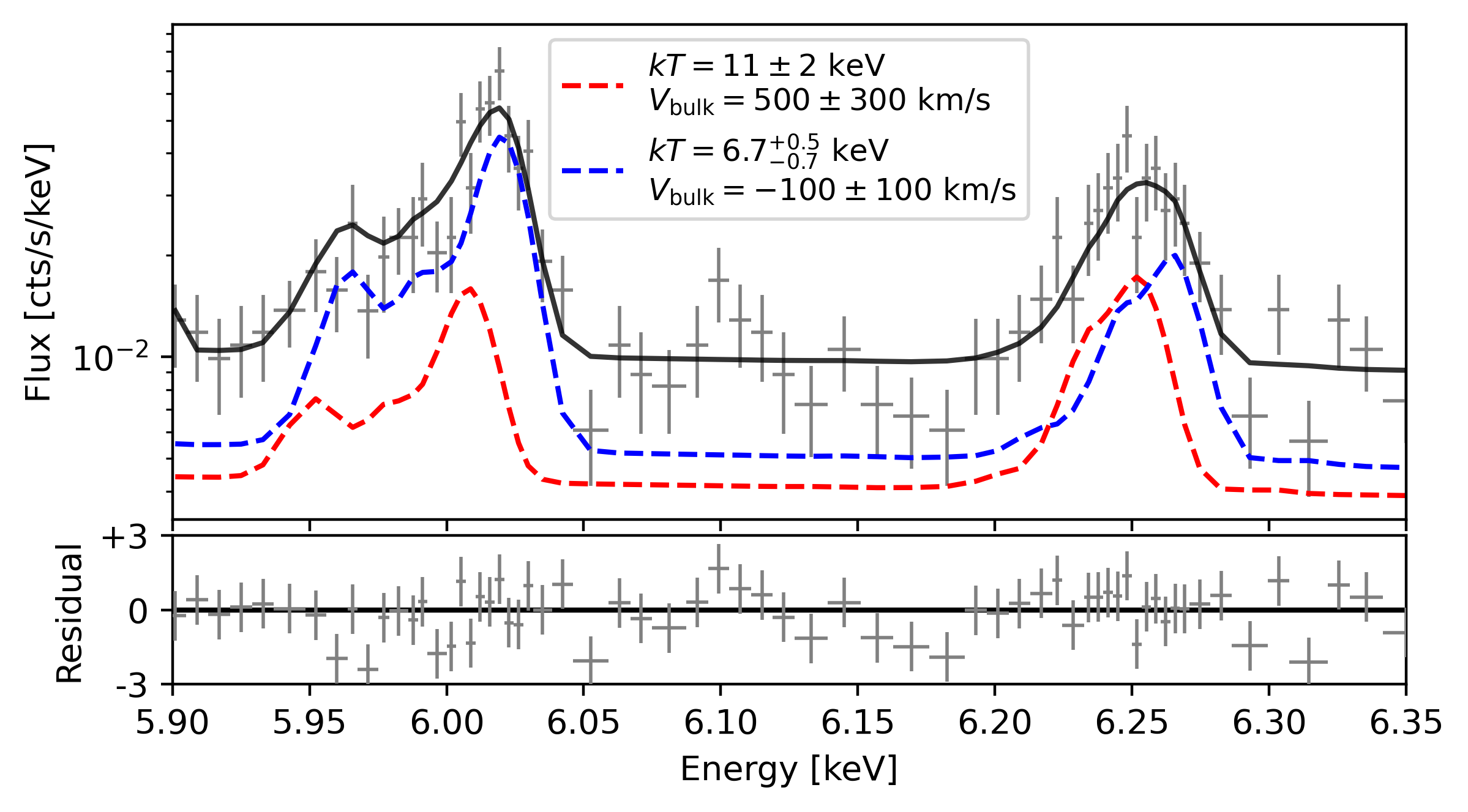}
    \caption{Full-array A2034 spectrum with best-fit two-component \texttt{bapec} model (black solid curve) and its components (colored dashed curves), with residuals plotted in the lower panel. Temperature and bulk velocity measurements are labeled. The spectrum is visually binned to $>3\sigma$ significance.
    }
    \label{fig:2comp}
\end{figure}

\end{appendix}

\end{document}